\documentclass[english,aps,prl,showpacs,amsmath,amssymb,twocolumn]{revtex4}
\usepackage{dcolumn}
\usepackage{bm}
\usepackage{epsf}
\makeatother
\begin{document}

\title{A closer look at the uncertainty relation of position and momentum}

\author{Thomas Sch\"urmann and Ingo Hoffmann}

\affiliation{J\"ulich Supercomputing Centre, J\"ulich Research Centre, 52425 J\"ulich, Germany}

\begin{abstract}
We consider particles prepared by a single slit diffraction experiment. For those particles the standard deviation $\sigma_p$ of the momentum is discussed. We find out that $\sigma_p=\infty$ is not an exception but a rather typical case. A necessary and sufficient condition for $\sigma_p<\infty$ is given. Finally, the inequality $\sigma_p\Delta x\geq \pi\hbar$ is derived and it is shown that this bound cannot be improved.
\end{abstract}

\pacs{03.65.Ta, 42.50.-p}

\maketitle

The diffraction of particles by a single slit has often been discussed as an illustration of Heisenberg's uncertainty relations and their role in the process of measurement. In the case of a single particle passing through a slit of width $\Delta x$ in a diaphragm of some experimental arrangement, the diffraction by the slit of the wave implies a spread in the momentum of the particle, which is greater the narrower the slit. This phenomenon is an example of the famous Heisenberg principle \cite{H27}\cite{H30}.

The most familiar formalization of the uncertainty principle is in terms of standard deviations \cite{H27}\cite{H30}\cite{K27}
\begin{eqnarray}\label{K}
\sigma_x\sigma_p\geq \hbar/2.
\end{eqnarray}

Here, the standard deviation of the position $\sigma_x$ is measured for a sample (beam) of particles initially prepared in a wavefunction $\psi$. Subsequently, the standard deviation of the momentum $\sigma_p$ is measured for a second sample of particles, which is also prepared in the state $\psi$. An experiment accepting this challenge has been performed in neutron interferometry and the results have been interpreted as an explicit verification of the uncertainty relation between position and momentum \cite{Kai83}. Nevertheless, the question has been discussed how to measure the mathematical terms of (\ref{K}). It has been argued that the neutron experiment does not measure the standard deviation $\sigma_x$ in position independently of certain ad hoc assumptions on the shape of the wavefunction such that the natural interpretation of this experiment might not be considered as a rigorous verification of the relation above \cite{Uff85}. Problems of this type have led to a number of different efforts towards a satisfactory formulation and proof of the uncertainty principle \cite{Far78}\cite{Pri83}\cite{Don89}.

An encouraging reformulation of (\ref{K}) has been proposed in terms of the {\it mean peak width} '$w$' and the {\it overall width} '${\cal W}$' of a wavefunction \cite{Uff85} (see also \cite{Don89} with regard to so-called $\epsilon$-concentrated functions). These measures are well defined and the type of problem mentioned above is partially solved by this approach. On the other hand, measures of this type are mostly related to suitable chosen (probability) numbers $N$ and $M$ (or concentration parameters in the case of \cite{Don89}), which are typically dependent on the experimental design and must initially be specified and prepared by the experimenter.

Alternatively the study of quantum information processing shows that
information theory might be suitable for the classification of quantum
uncertainty. Here, entropic uncertainty relations  provide a promising way
to express quantitatively the Heisenberg principle
\cite{Hi57}\cite{Be75}\cite{BiMy75}\cite{De83}\cite{P83}\cite{Bi84}\cite{MU88}\cite{SR98}\cite{Bi06}\cite{RS07}\cite{VS08}\cite{R08}.
 The first entropic relation corresponding to position and momentum was
proposed by Hirschman \cite{Hi57}. Namely he obtained an inequality for
position and momentum in terms of differential entropies and also
conjectured an improvement of his result. These conjectures have been
proved in \cite{Be75}\cite{BiMy75} using Beckner's formular for the
$(p,q)$-norm of the Fourier transform, while weaker results follow from the
Hausdorff-Young inequality.

In what follows, we will consider a different approach. For particles passing through a slit of width $\Delta x$, we  consider the diffraction of the incoming wave function $\psi$ as a preparation corresponding to the ordinary {\it von Neumann-L\"uders projection}. This approach is often applied in the actual experimental design of the uncertainty relation \cite{S69}\cite{Le69}\cite{Z03}. The advantage is that the localization of the particles is simply given by the width $\Delta x$ of the slit. Unfortunately this approach cannot be considered as a rigorous experimental test of expression (\ref{K}) because $\Delta x$ and $\sigma_x$ are quite different measures of localization. Furthermore, the standard deviation of the momentum, measured by the diffraction pattern at the screen, is not necessarily a finite number\cite{Be}\cite{UfHi}.

Therefore, in the following we first derive a necessary and sufficient condition to warrant finite standard deviations $\sigma_p$ of the momentum, given the particles have initially been prepared by a projection within $\Delta x$. Afterwards we will prove the inequality
\begin{eqnarray}\label{S}
\sigma_p\Delta x\geq \pi\hbar
\end{eqnarray}
and show that this bound cannot be further improved.\\

To do that, let us consider particles in one spatial dimension described by a wave function $\psi$ which is an element of the Hilbert space ${\cal H}=L^2(\mathbb{R})$, the space of square integrable functions on $\mathbb{R}$. The scalar product in Hilbert space will be denoted by angular brackets, that is to write $\langle \phi|\psi\rangle$ for the scalar product of two state vectors $\phi,\psi\in{\cal H}$. Accordingly, the norm of $\psi$ is given by $||\psi||\equiv \sqrt{\langle \psi|\psi\rangle}$.

Now, in one dimension the preparation by (ideal) diffraction is expressed by
\begin{eqnarray}\label{Neum}
\varphi(x)=\frac{\chi(x)\psi(x)}{||\chi(x)\psi(x)||},
\end{eqnarray}
while
\begin{eqnarray}\label{chi}
\chi(x) =\left\{ \begin{array}{r@{\quad\quad}l}
1 & \text{if }\quad |x|\leq \frac{\Delta x}{2},\\
0 & \text{otherwise} \label{abl}
\end{array} \right.
\end{eqnarray}
is the indicator function corresponding to the width of the slit. That is, the prepared state $\varphi(x)$ after the diffraction is a restriction of the initial state $\psi(x)$, typically centered around zero. In the following, we suppose that the overlap $\langle\chi|\psi\rangle\neq 0$. Notice that the computation of $\sigma_x$ is based on the function $\varphi(x)$ and we obtain $\sigma_x\leq\Delta x/2$ in any case.

The Fourier transform of $\varphi(x)$ gives the corresponding normalized state function $\hat\varphi(p)$ associated to the momentum of the particles after diffraction. In the definition of $\sigma_p$, the momentum probability density $|\hat\varphi(p)|^2$ is multiplied by the factor $p^2$, giving increasing weight to the distant parts of the probability distribution and the tails of the distribution often fall off too slowly for $\sigma_p$ to be finite. For instance, the diffraction by the slit in the case of a plane wave will imply a momentum distribution with infinite standard deviation $\sigma_p$ and (\ref{K}) is trivially satisfied.

A further important example is the diffraction of gaussian waves. Similar to the plane wave, the contribution for large $p$ after diffraction is given by a trigonometric term whose domain is restricted to oscillations within the unit interval, i.e. $p^2|\hat\varphi(p)|^2\propto\sin(\frac{\pi\Delta x}{h}\,p)^2$ for $|p|\to\infty$ and thus $\sigma_p=\infty$.

Now, let $f(x)$ be an integrable function on $\mathbb{R}$, its Fourier transform is the function $\hat f(p)$ on $\mathbb{R}$ defined by \cite{D70}
\begin{eqnarray}\label{four01}
\hat{f}(p) = \frac{1}{\sqrt{2\pi\hbar}}\int_{-\infty}^\infty e^{-\frac{i}{\hbar}p x}\,f(x)
\end{eqnarray}
We shall also occasionally write
\begin{eqnarray}\label{F01}
{\cal F}[f(x)] = \hat f(p)
\end{eqnarray}
for the Fourier transform of $f$. Without loss of generality, we now suppose the mean momentum of the particle after diffraction is zero. In this case the standard deviation of the momentum is
\begin{eqnarray}\label{std1}
\sigma_p^2 = ||p\,\hat\varphi(p)||^2
\end{eqnarray}
 and we have $\sigma_p<\infty$ if and only if the product $p\,\hat\varphi(p)$ is in the space of square integrable functions. If $\varphi(x)$ is continuous, piecewise smooth and the derivative $\varphi'(x)\in L^1(\mathbb{R})$, then we can write
\begin{eqnarray}\label{F2}
p \,\hat\varphi(p) = -i\hbar{\cal F}[\varphi'(x)].
\end{eqnarray}
After substitution into (\ref{std1}), we obtain
\begin{eqnarray}\label{std}
\sigma_p^2 = \hbar^2\,||{\cal F}[\varphi']||^2
=\hbar^2\,||\varphi'||^2,
\end{eqnarray}
while we have applied the theorem of Plancherel. Thus, $\sigma_p$ does exist if and only if $\varphi'(x)$ is in $L^2(\mathbb{R})$. Instead, it does not exist when the projection (\ref{Neum}) produces finite jumps at the edges of the slit. A formalization of that fact is given by the following:\\
\\
{\bf Lemma.} Let $\psi(x)\in {\cal H}$ be continuous and piecewise smooth. For every diffraction experiment corresponding to the projection rule (\ref{Neum}), the standard deviation $\sigma_p$ of the momentum does exist if and only if the derivative $\psi'(x)$ is square integrable on $D=[-\frac{\Delta x}{2},\frac{\Delta x}{2}]$ and $\psi(\pm\frac{\Delta x}{2})=0$.\\
\\
{\bf Proof.} According to (\ref{Neum}), the derivative of $\varphi(x)$ is formally given by the following expression
\begin{eqnarray}\label{deri}
\varphi'(x)&=& \frac{1}{||\chi(x)\psi(x)||}\,\Big[\,\chi(x)\psi'(x)
-\psi(\frac{\Delta x}{2})\,\delta(x-\frac{\Delta x}{2})\nonumber\\
&+&\psi(-\frac{\Delta x}{2})\,\delta(x+\frac{\Delta x}{2})\,\Big].
\end{eqnarray}
The Dirac distributions are not square integrable, that is,  $\psi'(x)\in L^2(D) \Leftrightarrow \varphi'(x)\in L^2(D)$ if and only if $\psi(\pm\frac{\Delta x}{2})=0$. Corresponding to (\ref{std}), $\sigma_p$ does exist if and only if $\psi'(x)\in L^2(D)$ and $\psi(\pm\frac{\Delta x}{2})=0$ are satisfied.   $\hfill\square$ \\

We are now in the position to obtain statements about the existence of $\sigma_p$ without explicit computation of the Fourier transform. Moreover, for all continuously differentiable states satisfying the boundary conditions of our lemma, we can now apply the Wirtinger inequality \cite{Wirt} by using (\ref{std}). After a few algebraic steps we obtain the useful relation
\begin{eqnarray}\label{K2}
\sigma_p\Delta x \geq \pi\hbar,
\end{eqnarray}
and this bound cannot be further improved. Actually, the equal sign in (\ref{K2}) is reached for the one-humped trigonometric function
\begin{eqnarray}\label{psi0}
\psi_0(x)=\sqrt{\frac{2}{\Delta x}}\;\cos(\pi x/\Delta x)
\end{eqnarray}
for $|x|\leq\Delta x/2$ and 0 otherwise. By computation of $\sigma_x$ we obtain the corresponding expression
\begin{eqnarray}\label{mu1}
\sigma_p\sigma_x=\sqrt{\frac{\pi^2-6}{3}}\;\frac{\hbar}{2}\approx 1.14\,\frac{\hbar}{2},
\end{eqnarray}
which is slightly above $\hbar/2$.

It should be mentioned that the value $\hbar/2$ can never be reached in (\ref{mu1}) because the projection (\ref{Neum}) does not produce gaussian densities with infinite support in position space, - the ordinary case of minimum uncertainty. However, for suitable truncated and shifted gaussian wavefunctions we might obtain the limit $\sigma_p\sigma_x\to\hbar/2$ if the width of the peak of such a wavefunction approaches zero. In this case, the left hand side in (\ref{K2}) approaches infinity because of $\sigma_p\to\infty$.

A similar but different inequality than (\ref{K2}) has been proven in \cite{Uff85} (eq.\,(21) therein) applying the mean peak width $w$ instead of $\Delta x$. However, this inequality has not been proven to be tight. Applied to the simple diffraction approach considered above, the inequality in \cite{Uff85} results in the (trivial) statement $\sigma_p\,w>0$, when $w=\Delta x$ with $M=1$.

\subsection{Conclusion}

A rigorous experimental verification of the uncertainty relation in single slit diffraction experiments requires a careful analysis of the measurement setup under consideration. In the ordinary case of plane waves, gaussian waves and many other types of generic wave functions with infinite support (see lemma), a straight approach is hard to establish because the standard deviation of the momentum does not exist. Finite values of $\sigma_p$ are only obtained for a special class of wave functions satisfying boundary conditions related to the width of the slit. That is, the most interesting case, i.e. $\sigma_p\sigma_x=\hbar/2$, can never be reached by any member of this class.

\newpage{}

\begin{thebibliography}{10}
\bibitem{Wirt} This inequality states that for any bounded interval $Q=[a,b]$ and any continuously differentiable $f(x)\in C^1(Q)$ with $f(a)=f(b)=0$: $\int_Q |f|^2 dx\leq [(b-a)^2/\pi^2]\int_Q |f'|^2 dx$. The constant $(b-a)^2/\pi^2$ cannot be improved \cite{D70}.
\end{thebibliography}
\end{document}